\documentclass{moriond}

\usepackage{amsmath}
\usepackage{amssymb}

\def\Journal#1#2#3#4{{#1} {\bf #2}, #3 (#4)}

\def\PRL{\em Phys. Rev. Lett.}
\def\PRD{{\em Phys. Rev.} D}

\def\be{\begin{equation}}
\def\ee{\end{equation}}
\def\bea{\begin{eqnarray}}
\def\eea{\end{eqnarray}}

\newcommand{\Photo}{}

\begin{document}
\vspace*{4cm}
\title{Searching for quasinormal modes from Binary Black Hole mergers}

\author{A.~Kr\'olak$^{1,2}$,
O.~Dorosh$^2$}

\address{$^1$Institute of Mathematics, Polish Academy of Sciences, 00-656 Warsaw, Poland\\
$^2$National Center for Nuclear Research, 05-400 Świerk-Otwock, Poland}

\maketitle\abstracts{We present a new method to search for gravitational waves from quasinormal modes in the ringdowns of the remnants of the mergers of the binary black hole systems. The method is based on maximum likelihood estimation. We derive a time-domain matched-filtering statistic that can be used to search for any number of modes in the data. The parameters of the modes can be estimated and the modes present in the data can be reconstructed. We perform Monte Carlo simulations of the method by injecting the quasinormal mode waveforms to noise. We analyze performance of the method for searches of quasinormal modes in the advanced detectors data like LIGO and Virgo, in the third generation of detectors like Einstein Telescope and Cosmic Explorer and in the space detector LISA data.
We analyze ringdown of publicly available GW190521 event and we compare our results with analysis by other methods.}

\section{Quasinormal modes}
\label{Sec:QNMs}

The gravitational-wave ringdown signal $s(t)$~\footnote{We restrict ourselves to prograde modes. See Sec. IIA of {\tt https://arxiv.org/abs/2107.05609} for discussion.} from a remnant black hole arising from a merger of a binary system 
can be expressed as
\be
\label{qnmsig}
s(t) = \sum_{lmn} [ A^c_{lmn}\, h^c_{lmn}(t)  + A^s_{lmn}\, h^s_{lmn}(t) ]
\ee
where $A^c_{lmn}$, $A^s_{lmn}$ are constant amplitudes and the time dependent functions $h^c_{lmn}(t)$, $h^s_{lmn}(t)$ are given by
\be
h^c_{lmn}(t) = \exp(-t/\tau_{lmn}) \cos(2\pi\, t\, f_{lmn}), \,\,\,\,\,\,\,
h^s_{lmn}(t) = \exp(-t/\tau_{lmn}) \sin(2\pi\, t\, f_{lmn})
\ee
where  $\tau_{lmn}$ and $f_{lmn}$  are  the damping time and frequency of the mode $(lmn)$ of the remnant, $(lm)$ are angular harmonic numbers and $n$ is the overtone number. By no-hair theorem the damping times and frequencies of the modes are determined by only two parameters of the remnant --- its mass $M$ and spin $a$.
The constant amplitudes $A^c_{lmn}$  and  $A^s_{lmn}$ depend in a complicated way on the initial conditions of the binary star merger.

\section{Detection statistic}
\label{Sec:QStat}

In this section we derive the matched-filtering statistic to detect and estimate parameters of an arbitrary 
number of quasinormal modes in the noise of the detector. It is necessary it identify more than one mode 
in order to test the {\em no-hair theorem}~\cite{Dreyer2004}.
We assume that the signal is additive, i.e. the data $x$ from a detector are give by
\be
x(t) = n(t) + s(t),
\ee
where $n(t)$ is the noise in the detector. 

Assuming that the noise in the detector is a Gaussian, stationary, and a zero-mean stochastic process
the log likelihood function reads
\be
\label{logL2}
\ln\Lambda = \big(x(t)\big|s(t)\big) - \frac{1}{2}\big(s(t)\big|s(t)\big),
\ee
where the scalar product $(\cdot|\cdot)$ is defined by
\be
(x|y) := 4\Re\int_0^\infty \frac{\tilde{x}(f)\, \tilde{y}^*(f)}{S_{h}(f)}\mathrm{d}f.
\ee
Here $S_{h}$ is the one-sided spectral density of the detector's noise
and tilde denotes Fourier transform.

To derive our statistic we first find the {\em maximum likelihood estimators} (MLEs) of the amplitude parameters by solving the following set of equations:
\be
\label{Amleqs2}
\frac{\partial \ln\Lambda}{\partial \mathcal{A}^c_k} = 0,  \,\,\,\,\,\,\,\,\,
\frac{\partial \ln\Lambda}{\partial \mathcal{A}^s_k} = 0, \quad k = 1,\ldots,K.
\ee
where $K$ is the number of modes and we have introduced a 3-dimensional multi-index $k = (lmn)$
going over all the modes. The solution of Eqs.\,(\ref{Amleqs2}) can be obtained in an explicitly analytic form.
\be
\hat{\mathcal{A}} = M^{-1} N,
\ee
where $\hat{\mathcal{A}}$ are MLEs of parameters $\mathcal{A}$ and the vector $N$ and matrix $M$ are given by
\be
N = \begin{pmatrix} (x|h^c_1)  \\
                    (x|h^s_1)  \\
\vdots \\ (x|h^c_K)  \\  (x|h^c_K)
\end{pmatrix},
\ee
\be
M = \begin{pmatrix} (h^c_1|h^c_1) & (h^c_1|h^s_1) & \ldots & (h^c_1|h^c_K) & (h^c_1|h^s_K) \\
                    (h^s_1|h^c_1) & (h^s_1|h^s_1) & \ldots & (h^s_1|h^c_K) & (h^s_1|h^s_K) \\  
\hdotsfor{5} \\
                    (h^c_K|h^c_1) & (h^c_K|h^s_1) & \ldots & (h^c_K|h^c_K) & (h^c_K|h^s_K) \\
                    (h^s_K|h^c_1) & (h^s_K|h^s_1) & \ldots & (h^s_K|h^c_K) & (h^s_K|h^s_K)
\end{pmatrix}.
\ee
After replacing in the log likelihood function \eqref{logL2}
the amplitudes $\mathcal{A}$ by their estimators $\hat{\mathcal{A}}$, 
we get the reduced likelihood function for the ringdown signal,
which we call the \emph{$\mathcal{Q}$-statistic},
\begin{align}
\label{Qstat}
\mathcal{Q}[x;\tau_k,f_k] := \ln\Lambda_I[x;\hat{\mathcal{A}},\tau_k,f_k]
= \frac{1}{2} N^T M^{-1} N.
\end{align}


We use two methods to analyze data with statistic $\mathcal{Q}$.
In the first --- {\em Kerr method}, we assume that the no-hair theorem holds, i.e. the damping times and frequencies are determined
by the mass $M$ and spin $a$ of the remnant black hole. 
Using the fits of $\tau_k$ and $f_k$ to  $M$ and $a$ \cite{Berti2006} we can obtain the statistic $\mathcal{Q}[x;M,a]$ for an arbitrary number of modes of the remnant with mass $M$ and spin $a$. 
In practice for the case of binary black hole mergers detected the parameters of the remnants are known with certain errors so we calculate the statistic $\mathcal{Q}[x;M,a]$ on a grid over the 2-dimensional space $(M,a)$ around the estimated values of $M$ and $a$.
In the second --- {\em agnostic method}, we calculate the statistic $\mathcal{Q}$ assuming that the data contains of a number $K$ of damped sinusoids with damping times $\tau_k$ and frequencies $f_k$. We then calculate the statistic  
$\mathcal{Q}[x;\tau_1,...,\tau_K,f_1,...,f_K]$ on a grid over $2 \times K$ dimensional parameter space.

\section{Monte Carlo simulations}
\label{Sec:MCs}
\begin{figure}
\begin{minipage}{0.5\linewidth}
\centerline{\includegraphics[width=\linewidth]{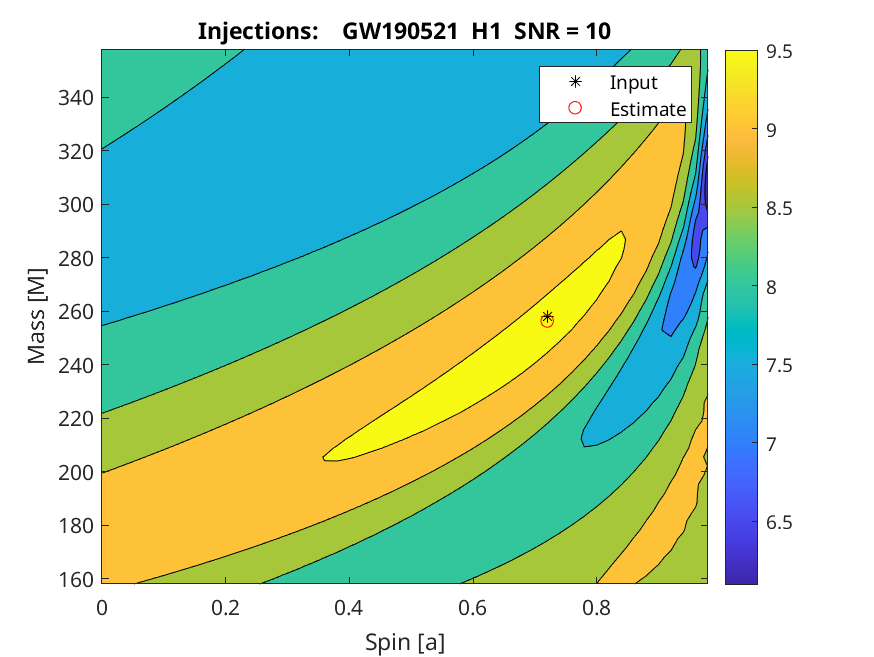}}
\end{minipage}
\hfill
\begin{minipage}{0.5\linewidth}
\centerline{\includegraphics[trim={5mm 0 5mm 0}, clip, width=\linewidth]{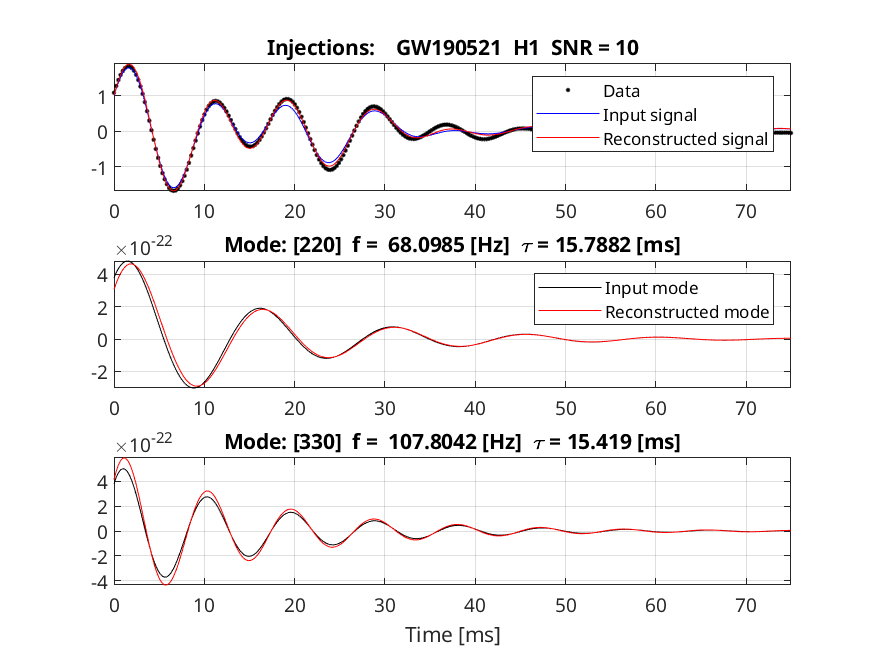}}
\end{minipage}
\caption[]{Ringdown signal consisting of the sum of two modes ($[2 2 0]$ and $[3 3 0]$ with M = $330\, M_\odot$, a = 0.86) added to LIGO H1 data. Left: SNR evaluated on $(M,a)$ parameter space. Right: the ringdown signal and the two components reconstructed using estimators of the parameters corresponding to the maximum of the $Q$-statistic.}
\label{fig:Kerr}
\end{figure}

To test our method we added artificial ringdown signals to the LIGO data~\cite{AdvLIGO} and we estimated 
parameters of the signal by evaluating the $Q$-statistic on a grid in the parameter space.
Estimators of the parameters are those for which the value of the $Q$-statistic is maximum.
An example is given in Figure \ref{fig:Kerr} where we have added a signal with parameters of the LIGO - Virgo (LVC) event GW190521~\cite{GW19} consisting of two modes. We searched for the signal using the Kerr method.  

We have performed Monte Carlo simulations by injecting signals consisting of two modes: $[2 2 0]$ (fundamental) and  $[3 3 0]$ with parameters of the GW190521 event to the LIGO Hanford detector (H1) data. We injected signals at random starting times 2 seconds after the GW190521 merger. We injected signals for an array of signal-to-noise ratios (SNRs) ranging from 5 to 20, and we performed 100 injection for each SNR. For the Kerr method  and for SNR $= 10$ the 1 standard deviations
were $\sigma_M = 28\, M_\odot$ and $\sigma_a = 0.075$  for estimation of mass and spin respectively.  For the agnostic method and SNR $= 10$ we obtained 1 standard deviations of $\sigma_{f_1} = 4$~Hz and $\sigma_{f_2} = 7.5$~Hz for the two frequencies $f_1$ and $f_2$ respectively and 
$\sigma_{\tau_1} = 9$~ms and $\sigma_{\tau_2} = 11$~ms for the two damping times $\tau_1$ and $\tau_2$. 

\begin{figure}
\hspace{-8mm}
\begin{minipage}{0.5\linewidth}
\centerline{\includegraphics[trim={5mm 0 5mm 0}, clip, width=\linewidth]{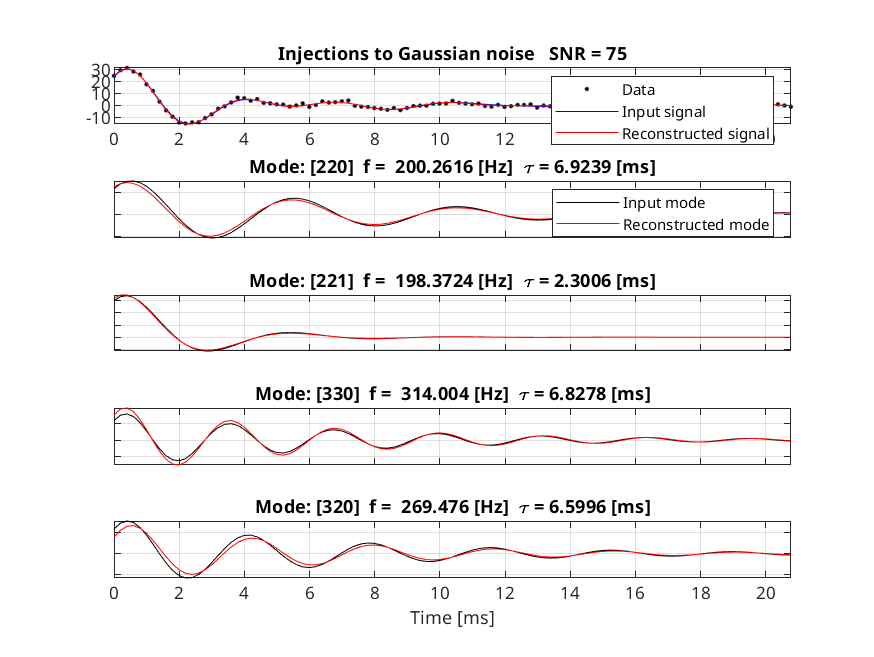}}
\end{minipage}
\hfill
\begin{minipage}{0.5\linewidth}
\centerline{\includegraphics[trim={1cm 0 1cm 0}, clip, width=\linewidth]{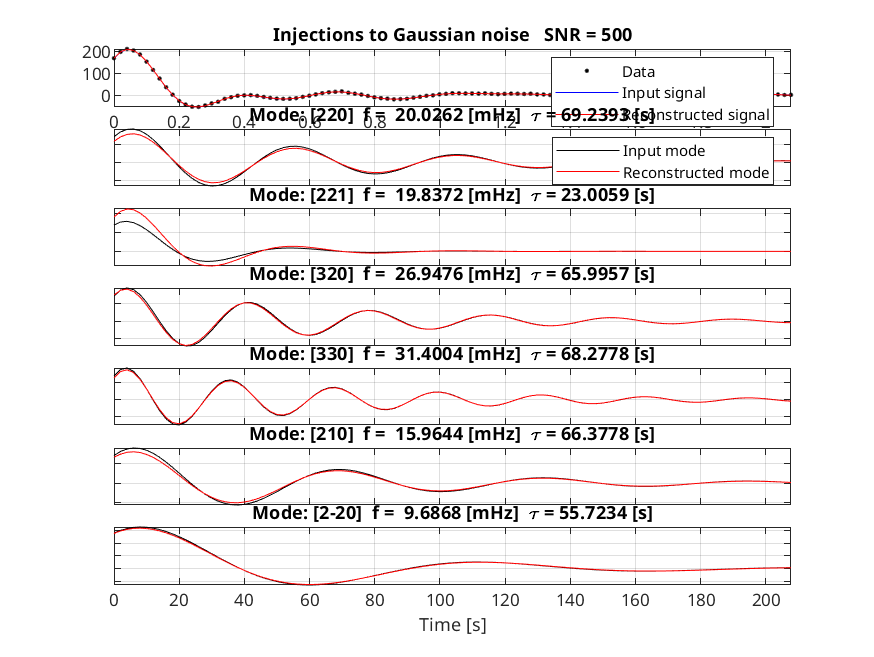}}
\end{minipage}
\caption[]{Left: 4 modes resolved for 3rd generation detector. SNR $=75$ (mass M = $100\, M_\odot$). 
Right: 6 modes resolved for LISA detector. SNR $=500$ (mass M = $10^6\, M_\odot$). In both cases spin a $= 0.85$.}
\label{fig:ET_CE_LISA}
\end{figure}

We also note that the waveform of quasinormal modes depends only on the spin $a$ which determines the number of cycles
of the signal and not on the mass $M$. Thus our pipeline which was primarily designed for analyzing data from currently operating ground based detectors will also be applicable to analyzing BBH ringdowns in the data of future 3rd generation detectors like Einstein Telescope~\cite{ET} and Cosmic Explorer~\cite{CE} and data from  space detector LISA~\cite{LISA}. The only difference is that much higher SNRs are expected and consequently more modes can simultaneously be resolved. As an example in 
Figure~\ref{fig:ET_CE_LISA} we have demonstrated that 4 modes can accurately be resolved for 3rd generation detector with 
SNR $= 75$,  and 6 modes for LISA detector where BBH ringdowns with SNRs of even 500 can be available. 

\section{Case study: GW190521 event ringdown}
\label{Sec:GW19}

We have applied our method to the case of GW190521 event. This is a high mass BBH merger with remnant mass of $258\, M_\odot$ (in the detector frame) and spin  $a = 0.72$~\cite{GW19}. For our analysis we have used the open LIGO-Virgo-KAGRA data available at

\begin{figure}
\begin{minipage}{0.50\linewidth}
\centerline{\includegraphics[width=\linewidth]{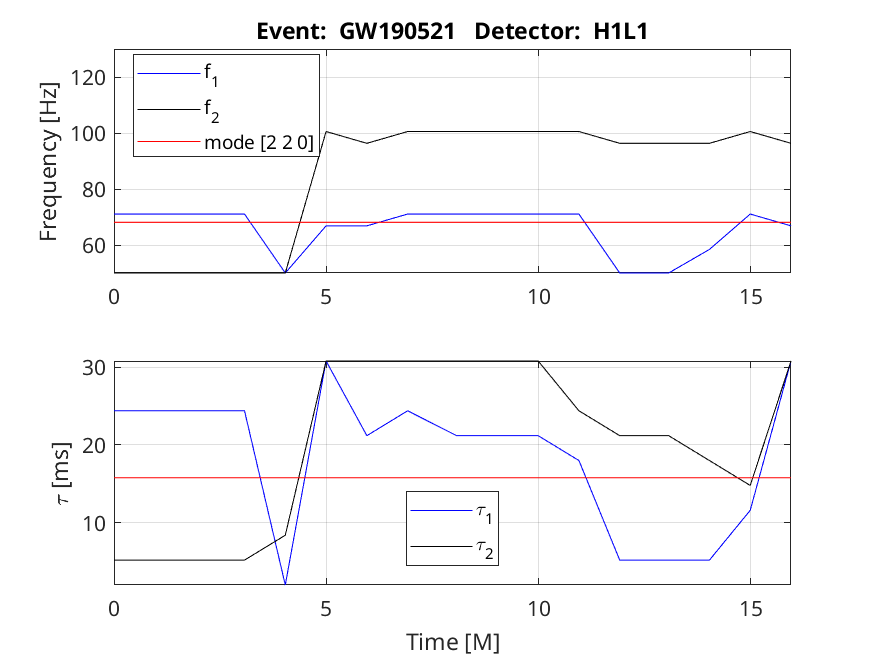}}
\end{minipage}
\hfill
\begin{minipage}{0.50\linewidth}
\centerline{\includegraphics[width=\linewidth]{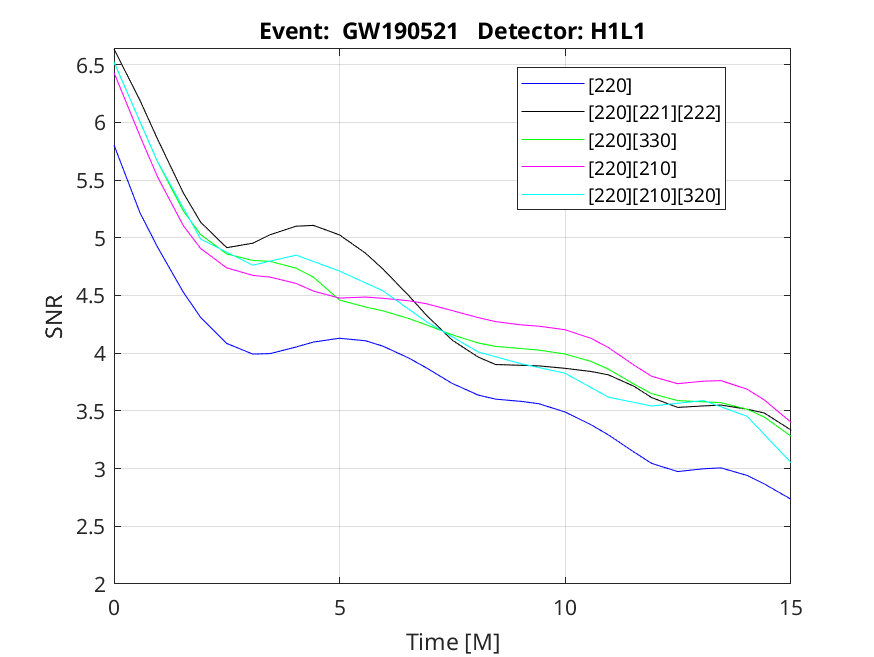}}
\end{minipage}
\caption[]{Left panel: estimates of frequencies $f_1$, $f_2$ and damping times $\tau_1$, $\tau_2$ for agnostic search of GW190521 event ringdown. The red horizontal lines correspond to estimates of mass and spin from analysis of the full signal~\cite{GW19}. Right panel: SNRs for fundamental mode and three combination of modes.
The $X$-axis on both panels is the start time $t_M$ of the analysis after the merger measured in total masses M of the system.}
\label{fig:GW190521}
\end{figure}

{\tt  https://gwosc.org/eventapi/html/GWTC-2.1-confident/GW190521/v4/}. For the analysis we preprocessed data by passing it through a narrow band filter with passband $[50 \,\, 130]$ Hz and whitening. As the starting point of the ringdown we chose the time $t_0$ at which the strain of the preprocessed data is maximum. We analyzed data from the two LIGO detectors. There is a rule of thumb that after the time 10 M (M is the total mass of the remnant measured in seconds) the perturbative approximation of the ringdown signal as the sum of damped sinusoids is valid \cite{Kamer}. Consequently we carried out our analysis for an array of ringdowns starting at times $t_M = n \times M$ after the time $t_0$ 
where $n$ was varied from 0 to 15. We have first carried out the agnostic search for two damped sinusoids in the ringdown data. The results are presented on the left panel of Figure~\ref{fig:GW190521}. The analysis found a frequency close to the frequency 68.1~Hz of the dominant quadrupole $[2 2 0]$ mode but also revealed another frequency around 100~Hz.
The nearby mode to this second frequency is $[3 3 0]$ with $f_{[3 3 0]} = 96.6$~Hz. However we need to keep in mind that there maybe unresolved frequencies near the frequency of the fundamental mode. We see that there is a mode $[2 1 0]$ with frequency 57.9~Hz close to the fundamental.
We have compared our findings with other analysis available in the literature. We found that the following quasimormal mode compositions were claimed for GW190521 ringdown: $([2 2 0],[2 2 1],[2 2 2])$ --- LVC~\cite{LVC}, $([2 2 0],[3 3 0])$ --- Capano et al.~\cite{Cap}, and $([2 2 0],[2 1 0]), ([2 2 0],[2 1 0],[3 2 0])$ ---  Siegel et al.~\cite{Sieg}. These findings overlap with our results.

Consequently we have carried out the Kerr analysis where we search for the ringdown signal composed of the four combinations of modes given above. For each mode combination we search for the maximum of the $Q$-statistic on a grid in mass-spin parameter space. We have also performed our analysis assuming that the ringdown consists only of the fundamental mode $[2 2 0]$. We have carried out the analysis for a number of starting times $t_M$ after the merger. The results are presented on the right panel of Figure~\ref{fig:GW190521}. 

We see that adding a second mode on top of the fundamental improves SNR by around 25\% at the time  $t_M = 10$. Concerning the additional mode we find that the combination $([2 2 0],[2 1 0])$ considered Siegel et al.~\cite{Sieg} gives the highest SNR. To summarize our analysis provides evidence for the presence of more than one quasinormal mode in the ringdown part of the GW190421 event.

Study of ringdown signals has been a subject of intensive research with data analysis methods concentrating mainly on the Bayesian approach~\cite{Cap,Sieg}. Here we propose an alternative maximum likelihood approach. Its advantage is reduction of the size of the parameter space by elimination of constant amplitudes from the analysis and concentrating on key parameters - mass and spin of the remnant. This leads in a considerable reduction in the computing time. The method can just provide a quick preliminary, nearly online search that can guide more refined Bayesian approaches.

\section*{Acknowledgments}

This research has made use of data or software obtained from the Gravitational Wave Open Science Center ({\tt gwosc.org}),
a service of LIGO Laboratory, the LIGO Scientific Collaboration, the Virgo Collaboration, and KAGRA.
LIGO Laboratory and Advanced LIGO are funded by the United States National Science Foundation (NSF)
as well as the Science and Technology Facilities Council (STFC) of the United Kingdom,
the Max-Planck-Society (MPS), and the State of Niedersachsen/Germany for support of
the construction of Advanced LIGO and construction and operation of the GEO600 detector.
Additional support for Advanced LIGO was provided by the Australian Research Council.
Virgo is funded, through the European Gravitational Observatory (EGO),
by the French Centre National de Recherche Scientifique (CNRS),
the Italian Istituto Nazionale di Fisica Nucleare (INFN) and the Dutch Nikhef,
with contributions by institutions from Belgium, Germany, Greece, Hungary, Ireland, Japan, Monaco, Poland, Portugal, Spain.
KAGRA is supported by the Ministry of Education, Culture, Sports, Science and Technology (MEXT),
Japan Society for the Promotion of Science (JSPS) in Japan;
National Research Foundation (NRF) and Ministry of Science and ICT (MSIT) in Korea;
Academia Sinica (AS) and National Science and Technology Council (NSTC) in Taiwan.
We would also like to thank Harrison Siegel for helpful discussions.
The work was supported by the Polish National Science Centre Grant No.\ 2023/49/B/ST9/02777.

\section*{References}
\bibliography{moriond}


\end{document}